\documentclass[twocolumn,english,superscriptaddress]{revtex4-1}
\usepackage[latin9]{inputenc}
\usepackage{babel}
\usepackage{units}
\usepackage{amssymb}
\usepackage{graphicx}
\usepackage{natbib}
\usepackage[unicode=true,pdfusetitle,
 bookmarks=true,bookmarksnumbered=false,bookmarksopen=false,
 breaklinks=false,pdfborder={0 0 1},backref=false,colorlinks=false]
 {hyperref}

\makeatletter

\renewcommand{\cite}{\citep}
\renewcommand{\citet}{\citep}

\makeatother

\begin{document}
\title{Fast tunable high Q-factor superconducting microwave
resonators}
\author{Sumedh Mahashabde}
\affiliation{Department of Microtechnology and Nanoscience MC2, Chalmers University
of Technology, SE-41296 Goteborg, Sweden}
\author{Ernst Otto}
\affiliation{Department of Microtechnology and Nanoscience MC2, Chalmers University
of Technology, SE-41296 Goteborg, Sweden}
\author{Domenico Montemurro}
\affiliation{Department of Microtechnology and Nanoscience MC2, Chalmers University
of Technology, SE-41296 Goteborg, Sweden}
\author{Sebastian de Graaf}
\affiliation{National Physical Laboratory, Hampton Road, Teddington, TW11 0LW,
United Kingdom}
\author{Sergey Kubatkin}
\affiliation{Department of Microtechnology and Nanoscience MC2, Chalmers University
of Technology, SE-41296 Goteborg, Sweden}
\author{Andrey Danilov}
\email{andrey.danilov@chalmers.se}
\affiliation{Department of Microtechnology and Nanoscience MC2, Chalmers University
of Technology, SE-41296 Goteborg, Sweden}
\begin{abstract}
We present fast tunable superconducting microwave resonators fabricated from
planar NbN on a sapphire substrate. The $3\lambda/4$ wavelength resonators
are tuning fork shaped and tuned by passing a dc current which controls the
kinetic inductance of the tuning fork prongs. The $\lambda/4$ section from
the open end operates as an integrated impedance converter which creates a
nearly perfect short for microwave currents at the dc terminal coupling
points, thus preventing microwave energy leakage through the dc lines. We
measure an internal quality factor $Q_{\rm int}>10{^{5}}$ over the entire
tuning range. We demonstrate a tuning range of $> 3\%$ and tuning
response times as short as 20 ns for the maximum achievable detuning. Due to
the quasi-fractal design, the resonators are resilient to magnetic fields of
up to 0.5 T.

\end{abstract}
\maketitle

\section{{\large{}Introduction}}

Superconducting microwave resonators are versatile devices with applications
ranging from signal conditioning in Purcell filters \citet{reed2010fast} and
parametric amplifiers
\citet{wustmann2013parametric,simoen2015characterization} to sensing
applications such as kinetic inductance detectors \citet{day2003broadband}
and near field scanning microwave microscopes
\citet{de2013near,de2015coherent,geaney2019near}. In particular, in the
rapidly developing field of quantum computing superconducting resonators are
used for qubit readout, on demand storage/release \citet{pierre2014storage}
and routing of single microwave photons
\citet{yin2013catch,hoi2011demonstration}, and deployed as qubit
communication buses \citet{wendin2017quantum}. Furthermore, future progress
in quantum computing calls for identification and elimination of material
defects, which spoil coherence in quantum circuits
\citet{paladino20141,muller2019towards}. This urgent need stimulated the
development of planar Electron Spin Resonance (ESR) spectrometers with
sub-femtomole sensitivity \citet{de2017direct,de2018suppression} to
environmental spins. The frequency tuning functionality either greatly
benefits (filters, sensors, spectrometers) or is instrumental for (parametric
amplifiers, single photon sources) all the above applications.

The base frequency of any microwave resonator is defined by its geometry and
can therefore be adjusted by varying the geometrical parameters. In practice,
purely mechanical designs are bulky, challenging to implement
\citet{kim2011thin} and do not provide fast tuning.

Conveniently, superconducting design elements provide an alternative option:
any superconductor possesses kinetic inductance (KI), which can be tuned with
either external magnetic field or dc bias current. The first approach does
not require any tweaks to a standard coplanar resonator design and was
presented already a decade ago by Healey et al. \citet{healey2008magnetic}.
However, sweeping the external magnetic field is also a rather slow solution;
the kinetic inductance itself can respond in sub-nanosecond time (the
instantaneous KI response is, for example, exploited in traveling wave
parametric amplifiers \citet{eom2012wideband,chaudhuri2017broadband} and comb
generators \citet{erickson2014frequency}).

A standard way to achieve fast frequency control via KI tuning is by
integrating a pair of Josephson junction elements in the form of a
Superconducting Quantum Interference Device (SQUID) loop in the resonator
design. The highly nonlinear loop inductance can be controlled with a
magnetic flux generated by a dedicated current control line, such that no
external field is needed. Devices of this type achieve tuning speeds faster
than the photon lifetime $Q/f_{0}$ \citet{sandberg2008tuning}, and recently a
frequency tuning time less than $\sim 1$ ns was demonstrated
\citet{wang2013quantum}. Today, SQUID-tunable resonators have become a common
element in circuit QED experiments
\citet{palacios2008tunable,osborn2007frequency,kennedy2019tunable}.
Unfortunately, the insertion of Josephson elements degrades the resonance
quality: first designs demonstrated $Q \sim 3000$, and in best-ever devices
$Q$ still does not exceed 35000 \citet{svensson2018microwave}. This is
significantly lower than standard non-tunable coplanar resonators, where $Q
\sim 10^{6}${} is now typical \citet{megrant2012planar}.

As an alternative to SQUIDs as lumped KI elements, one can exploit the
kinetic inductance of the superconducting film itself. This approach does not
introduce the extra internal dissipation associated with SQUIDs, but
generally does increase radiation losses: the microwave energy leaks from the
cavity through the dc terminals. An obvious way to minimize the losses is to
couple the dc leads at the nodes of the microwave voltage, but, due to
fabrication tolerances, one does not reach $Q > 1000$ in a practical device.
The leakage can be further suppressed by supplying the bias current through
low pass filters. The simplest of such filters, capacitors, provide a shunt to
ground for microwaves, thus effectively decouplindg dc lines. With this
design, internal quality factors up to $10^{4}$ in tunable coplanar
resonators were demonstrated in \citet{bosman2015broadband}. The maximum $Q$
is limited by two factors: the filter capacitor cannot be made larger than
few tens of pF because of self-resonances, and the capacitor is non-ideal
because of the finite loss tangent of the dielectric layer. The dielectric
losses were the the limiting factor in \citet{bosman2015broadband}. With a
better dielectric such as aluminium oxide deposited by atomic layer
deposition (ALD), $Q$ factors up to $10^{4}$ in tunable microstrip resonators
have been reached \citet{adamyan2016tunable}. Another simple filter type, an
inductive-resistive (LR) filter, was reported in \citet{vissers2015frequency}. With
the choke inductance $L_{\rm F}=2\thinspace$ nH and resistance $R_{\rm
F}=0.04\thinspace\Omega$, $Q$-factors up to $10^{5}$ were reported. However,
this solution involves the trade-off between the quality factor and the
maximum tuning rate: the $Q_{\rm int}$ is limited by the losses in the
dissipative $R_{\rm F}$, therefore $R_{\rm F}$ has to be kept small, and a
high $L_{\rm F}$ is needed to effectively isolate $R_{\rm F}$; both
requirements slow down the response time $\tau=L_{\rm F}/R_{\rm
F}=50\thinspace$ ns. If few-nanosecond tuning time is required, the
achievable $Q$-factor would not exceed that of SQUID-tunable resonators.

Here, we present a novel resonator design which incorporates an impedance
converter as an integral part of the resonator circuit. The converter
provides a nearly perfect short circuit for microwave currents at the dc
terminals insertion points thus preventing an energy leakage into terminals.
By design, this solution imposes no limits on the frequency tuning rate and
we report dissipative $Q$-factors above one million at high photon
populations and up to $2 \times 10^{5}$ at single photon level, approaching
that of standard non-tunable superconducting coplanar resonators.

\begin{figure}
\includegraphics{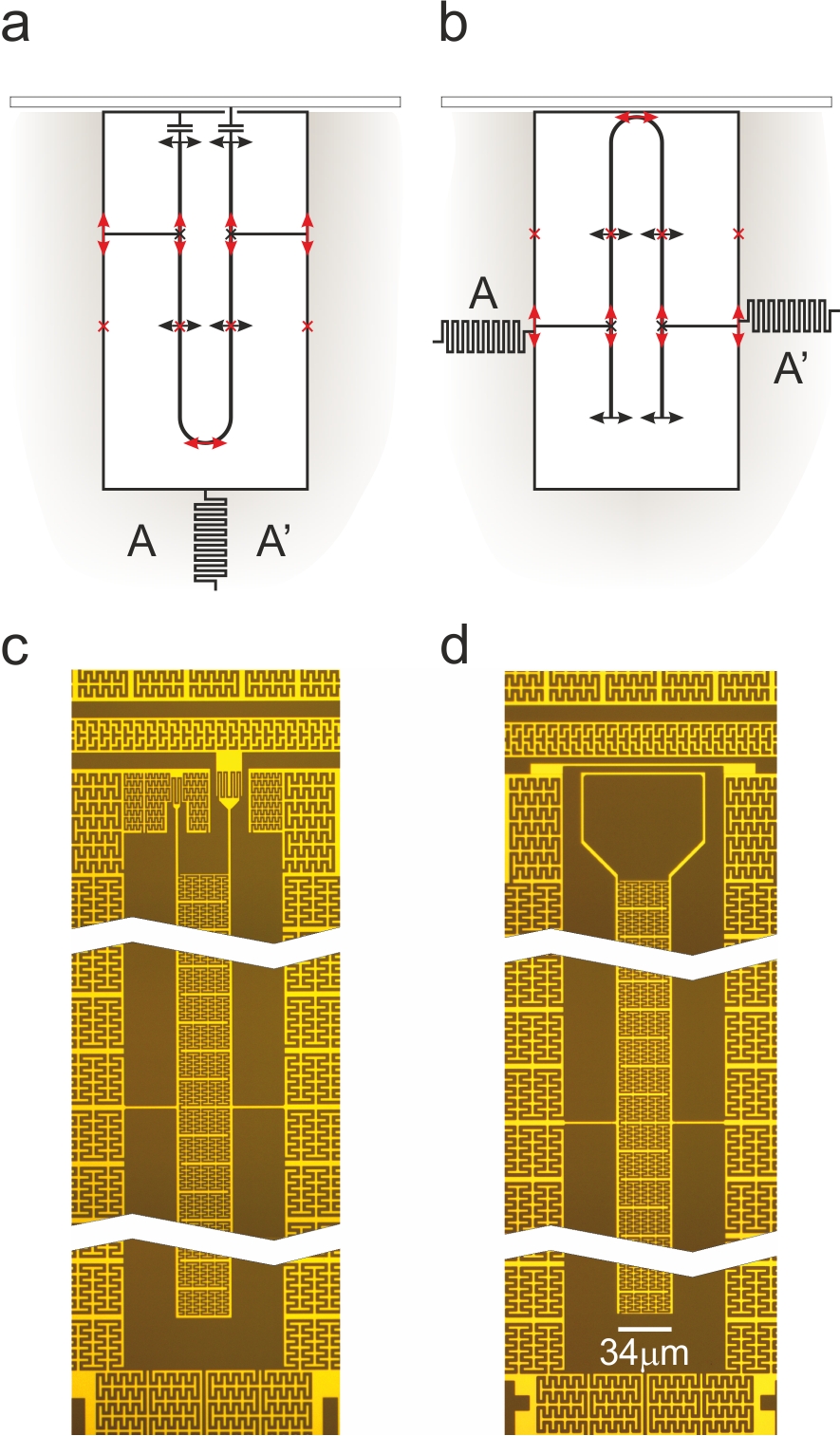}

\caption{\textit{Design concept and optical images.} (a) and (b) show a
cartoon representation of a capacitively and inductively coupled tuning fork
resonator designs. Black and red arrows show the positions of voltage and
current maximums respectively. Black and red crosses mark the positions of
voltage and current nodes. A and A' are the two galvanically split ground
half-planes. Corresponding optical images of the resonators
are shown in (c) and (d).}

\label{Fig 1}
\end{figure}

\begin{figure*}
\includegraphics[width=150mm]{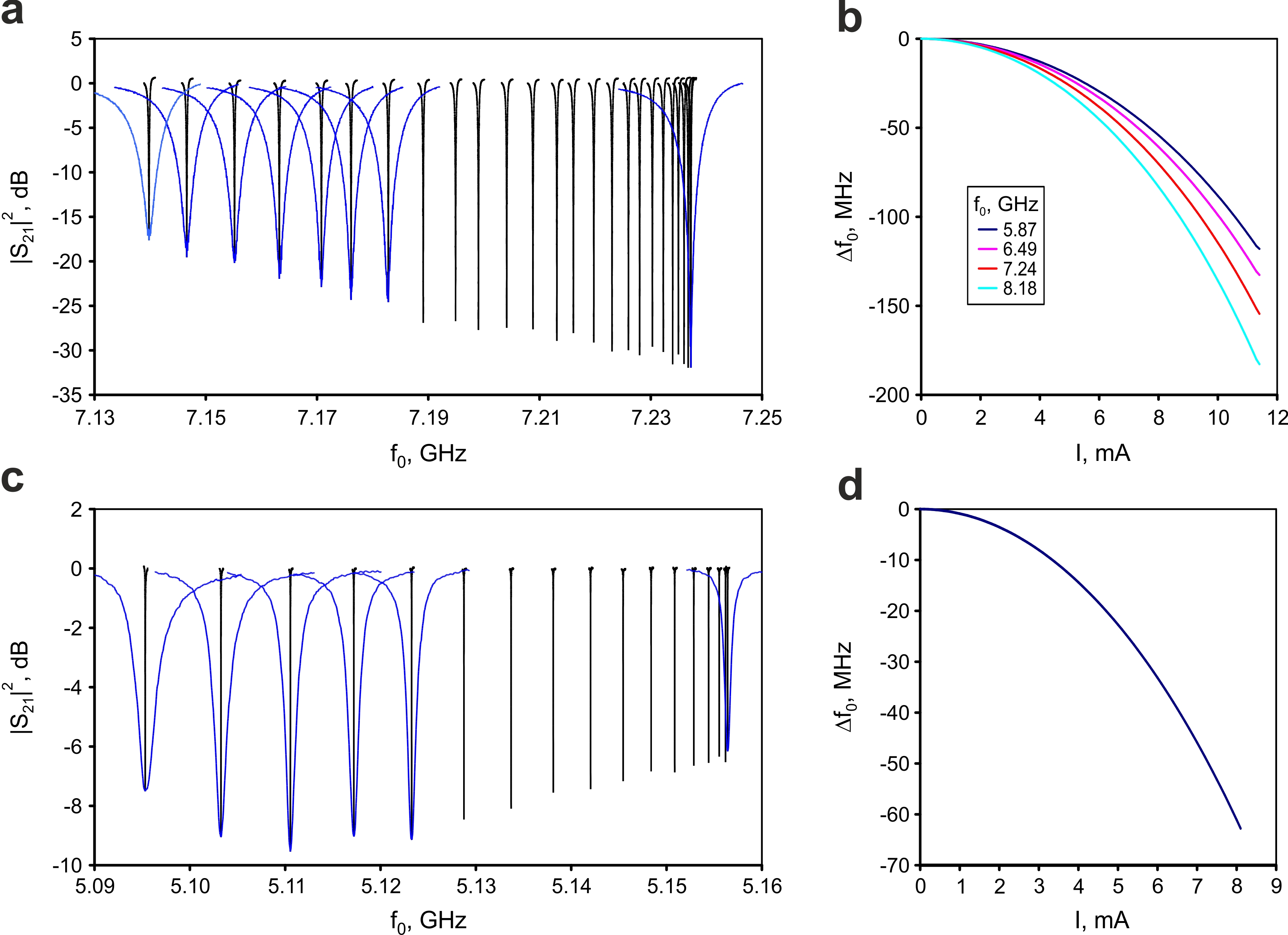}

\caption{\textit{Frequency tuning with dc bias current.} (a,c) A set of
$S_{21}$ transmission resonances taken at different bias currents. The
measured curves are shown in black; some curves (shown in blue) have been
stretched in frequency (25 times for (a) and 100 times for (c)) to
demonstrate the shape of the resonances. (b,d) Frequency shift versus the
bias current. The (a,b) plots are for capacitively coupled and
(c,d) for inductively coupled designs.}

\label{Fig 2}
\end{figure*}

\section{{\large{}Methods}}

The resonator design is sketched in Fig. \ref{Fig 1}(a, b). Conceptually, the
resonator core is an electromagnetic analog of the mechanical tuning fork.
The coupling to a readout coplanar waveguide transmission line is provided
either via an interdigitaited capacitor (a) or by an inductive loop (b). As
it will be demonstrated below, the resulting device performance does not
depend on the coupling scheme, so any solution can be freely chosen for a
specific target application. The detailed design can be seen in optical
images presented in Fig. \ref{Fig 1}(c, d): the two tuning fork prongs are
designed to be 2 $\mu$m in width. The prongs are inductive elements which are
coupled via an interdigitated capacitor formed by a 3rd order quasi-fractal
pattern. The role of the quasi-fractal design will be discussed in detail
below; in summary it is a descendent of non-tunable resonators presented
earlier in \citet{graaf2012magnetic,de2014galvanically}.

We exploit the $3\lambda/4$ mode of the tuning fork. The current/voltage
nodes/antinodes of this mode are depicted in Figs. 1(a,b): the $3\lambda/4$
mode supports a voltage node at a distance of $\lambda/4$ away from the open
end of the resonator. This is the point on the prongs where we inject a dc
current via a galvanic connection to the ground plane. The bias current thus
flows along the closed U-turn section of the fork and controls the kinetic
inductance of the resonator's $\lambda/2$ section, allowing the resonance
frequency to be tuned. To allow for the dc bias, the ground plane is split
into two sections A and A', galvanically separated by a large interdigitated
capacitor(s) (sketched as meander(s) in Fig. \ref{Fig 1}(a, b)), that present
negligible impedance for microwave currents ({\it cf.}
\citet{de2014galvanically} for further details). The finite fabrication
tolerances result in some residual voltage at the nominal nodal points; the
dc terminal lines, if present, couple to this voltage as antennas, resulting
in radiation losses. In contrary, the split ground plane solution eliminates
radiating dc lines; the residual currents instead circulate across the
splitting capacitor. Most crucially, the $\lambda/4$ segment (from the bias
injection points till the open end) presents an impedance converter, which
translates an infinite open end impedance into essentially zero impedance
between the dc coupling points. The effective microwave short zeroes the
residual voltage and dramatically reduces radiation losses.

Another dissipation channel appears if the resonator mode couples to some
parasitic resonance supported by the ground plane or the enclosure. For a
fixed-frequency resonator the odds that the resonance frequency will match
some parasitic resonance are relatively low, and, if such unlikely collision
does take place, the problem could be eliminated by a few extra wire-bonds
across the chip which will shift the parasitic resonance in frequency. For
tunable resonators this simple solution obviously does not apply: all ground
plane resonances present within the full tuning range (100-200 MHz for the
presented design) should be either eliminated or decoupled from the
resonator. In the presented resonators this challenge is conjointly addressed
by a set of design solutions listed below. Firstly, the tuning fork geometry
ensures that the microwave mode is localized in-between the prongs. Secondly,
the fractalized prong-to-prong capacitor provides per unit length capacitance
much higher than that of a regular coplanar line; this translates into a very
slow phase propagation velocity ($\sim 4\%$ of the speed of light) and a
short resonator length of about 1.5 mm for resonance frequency of 6 GHz. The
transverse dimension is also rather compact: the prong-to-prong distance is
34 $\mu$m only. As the coupling to environmental resonances is proportional
to the square of the dipole moment, our resonators are much more resilient to
parasitics than the regular coplanar lines. Finally, elimination of dedicated
dc lines and the simplest ground plane topology minimizes the spectral
density of parasitics. In practice, we achieve above 75\% yield: out of four
resonances on a chip no more than one typically suffers from collision with
parasitics within the full tuning range.

The resonators described here were fabricated from 140 nm NbN film deposited
on a 2-inch sapphire substrate. The desired structure is patterned with electron
beam lithography and etched in a $Ar:Cl_{2}$ reactive ion plasma. The
fabrication procedure is broadly similar to \citet{graaf2012magnetic}, where
the $Ar:Cl_{2}$ reactive ion etching was used to ensure sharp sidewalls and
to prevent lateral under-resist etching \citet{niepcenanowire2018}. $Au$
bonding pads were patterned and deposited in the next fabrication layer.
Finally, with one extra exposure, the resonator (but not the ground) was
thinned down to the desired thickness (50 nm) using the same $Ar:Cl_{2}$
reactive ion etch process. This thickness yields a sheet kinetic inductance
of $\sim 4$ pH/$\square$, which places the resonance frequency in the 4-8 GHz
band. The ground plane thickness was left unchanged with sheet kinetic
inductance $\sim 1$ pH/$\square$. This ensures that the typical frequencies
of the ground plane resonances are placed above 8 GHz. Each 2-inch wafer yields
12 chips of size $10 \times 9$ mm$^{2}$. The chips with the capacitively
coupled design had 4 resonators on the chip, while for the inductively
coupled design there was only
a single resonator per chip.

\begin{figure}
\includegraphics{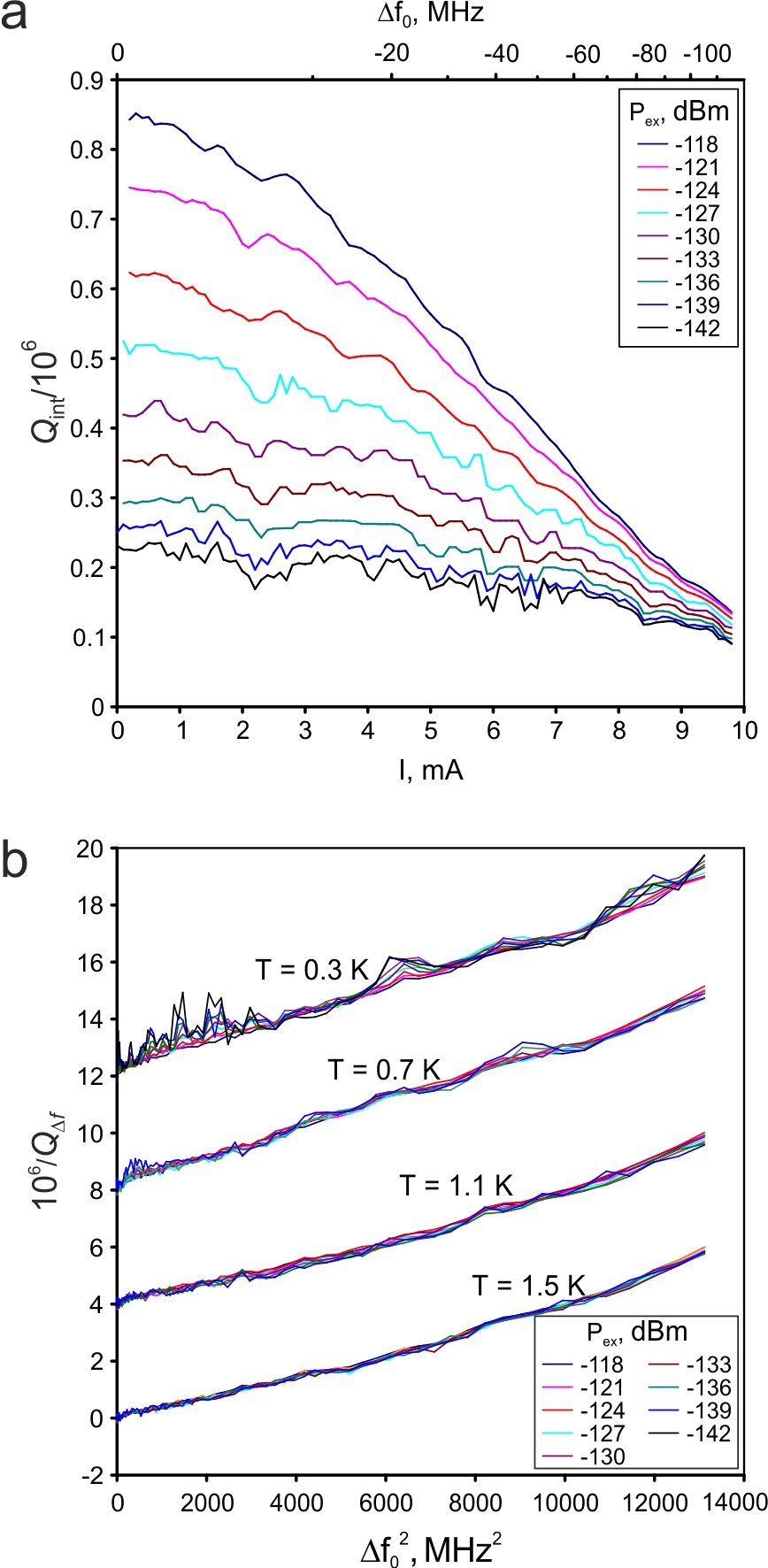}

\caption{\textit{Q-factor vs. frequency tuning.} (a) As measured. (b)
Tuning-related dissipation part $1/Q_{\Delta f}$. The plots are presented vs.
a square of the frequency shift $\Delta f$ to illustrate the linear in
$(\Delta f_{0})^{2}$ dependence. The plots taken at different temperatures
are offset for clarity (same offset for all excitation
powers $P_{ex}$).}

\label{Fig 3}
\end{figure}

Based on the normal state sheet resistance of the film, the kinetic
inductance can be estimated as $L_{\rm k}=\displaystyle{\hbar R_{\rm
N}}/{\pi\Delta_{0}}$, where $\hbar$ is the reduced Planck constant, $R_{\rm
N}$ is the sheet resistance, and $\Delta_{0}$ is the superconducting energy
gap at 0 K. This corresponds to a kinetic inductance of 4 pH/$\square$ for
the 50 nm thick film. This formula is approximate but allows an estimation of
$L_{\rm k}$ which can be used to model the resonator design with the Sonnet
Electromagnetic Simulation Software \citet{sonnet}. From the simulation, the
position of the voltage node in the 3$\lambda$/4 mode is obtained in
order to connect the dc current leads at the optimal place.

\begin{figure}
\includegraphics{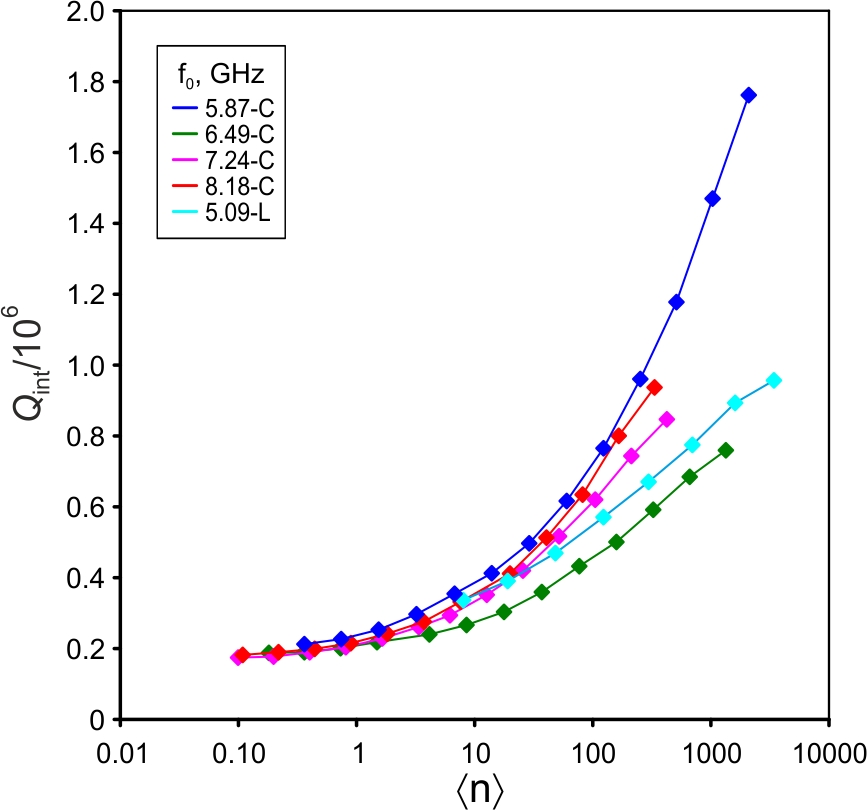}

\caption{\textit{Internal Q-factor for different photon occupations $\langle
n \rangle$.} In the legend C and L stand for capacitively and inductively
coupled resonators. Measured at 0.3 K.}

\label{Fig 4}
\end{figure}

\begin{figure*}[t]
\includegraphics[width=17.8cm]{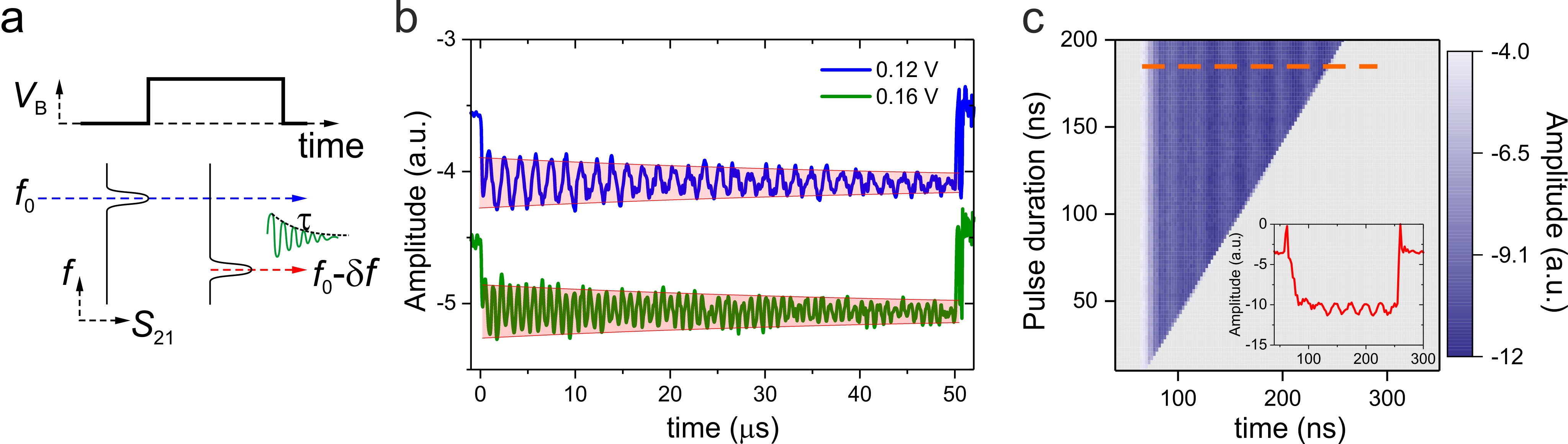}

\caption{\textit{Characterization of the tuning response time.} (a) Bias
control pulse (top) and dynamic resonator response. (b) Example of the
measured homodyne detector output response to two detuning pulses (offset for
clarity). The base resonator frequency is 6.15 GHz; the blue plot is for 0.6
MHz and the green plot is for 1.2 MHz detuning. The exponential energy decay
with characteristic time $\tau=Q/\pi f_{0}=50$ ${\rm {\mu}}s$ is indicated by
the envelopes. (c) Resonator response to detuning pulses of the same
amplitude (corresponding to 40 MHz detuning) but different durations. Inset:
cross-section of the main color plot indicated with the dashed line.}

\label{Fig 5}
\end{figure*}

\section{{\large{}Results and Discussion}}

The fabricated devices were initially characterised in a 2 K LHe cryostat. A
vector network analyser (VNA) was used to record the forward transmission
($S_{21}$) of the device; the measured $S_{21}$ data is shown in Fig.
\ref{Fig 2} (a,c). The resonance frequency smoothly changes as a function of
bias current following a parabolic dependence ($cf.$ Fig. \ref{Fig 2}(b,d))
$L_{\rm k}(I)\approx L_{\rm k}(0)(1+(\displaystyle{I_{\rm
dc}}/{I_{*}})^{2})$, where $L_{\rm k}(0)$ is the kinetic inductance at zero
current, $I_{\rm dc}$ is the bias current and $I_{*}$ is the nonlinearity
parameter. The $I_{*}$ extracted from a parabolic fit to Fig. \ref{Fig
2}(b,d) gives $I_{*}\approx55$ mA such that $\displaystyle{I_{*}}/{I_{\rm
c}}\approx5$, where $I_{\rm c}$ is the critical current of the resonator.
These parameters can be to some extent tuned by varying the NbN film
deposition conditions such as the substrate temperature, $N_{2}$ partial
pressure and the flow rate.

For further characterization, we used a single shot $^{3}\rm He$ cryostat
with a base temperature of 0.3 K. $S_{21}$ scans were recorded at different
temperatures, microwave excitation power levels, and dc tuning currents. From
the recorded $S_{21}$ scans, the internal and coupling quality factors were
extracted by fitting the data to the model provided in
\citet{probst2015efficient}; the aggregated results are shown
in Fig. \ref{Fig 3}.

Fig. \ref{Fig 3}(a) presents the internal quality factor as a function of dc
tuning for different excitation powers for the measurements taken at 0.3 K.
At zero tuning, $Q_{\rm int}$ depends on the excitation power (the power
dependence is addressed later). Tuning the resonator frequency induces some
excessive dissipation and $Q_{\rm int}$ decreases. Once $Q_{\rm int}$
approaches $\sim10^{5}$, the tuning-related part of dissipation starts to
dominate and the total $Q$-factor does not depend on the excitation power
anymore. Similar measurements were also performed at 0.7 K, 1.1 K and 1.5 K.
Quantifying the tuning related quality factor as $1/Q_{\Delta \rm f}=1/Q_{\rm
int}(I)-1/Q_{\rm int}(0)$, we compose a plot shown in Fig. \ref{Fig 3}(b),
where $1/Q_{\Delta \rm f}$ is presented as a function of $\Delta f_{0}^{2}$.
The fact that $1/Q_{\Delta \rm f}$ does not depend on the excitation power or
on the original (non-tuned) $Q_{\rm int}$ clearly indicates that tuning
related dissipation is of purely radiative nature. In fact, the quadratic
dependence on frequency tuning $\Delta f_{0}$ is what should be expected for
radiative losses in our design: the dc current only affects the inductance of
the $\lambda/2$ part of the resonator at the closed end (see Fig. \ref{Fig
1}(a,b)) and does not affect the inductance of the $\lambda/4$ part at the
open end. As a result, as the resonance frequency is tuned, the position of
the voltage nodes is slightly shifted away from the dc terminal coupling
points. The voltage node shift is proportional to $\Delta f_{0}$, and thus
the excessive radiation to $\Delta f_{0}^{2}$. For applications which do not
require extreme tuning time as short as $1/f_{0}$, the residual radiation
losses can further be suppressed by replacing direct prong-to-ground
links with simple low pass filters.

An interesting feature in Fig. \ref{Fig 3} are ripples appearing in $Q_{\rm
int}$ plots at the near single-photon excitation powers and the lowest
temperatures. These ripples are reproducible on short (seconds-to-minutes)
timescales and non-reproducible on longer timescales (a couple of hours). We
attribute these fluctuations in $Q_{\rm int}$ to interaction with an ensemble
of charged two-level systems (TLS), as previously reported in
\citet{brehm2017transmission}. At high temperatures/excitation powers, the
TLS are saturated and the observed ripples are smeared. Also, as the energies
of individual fluctuators are slowly varying in time
\citet{muller2019towards}, the ripple pattern is not reproducible on a longer
time scale.

Figure \ref{Fig 4} presents the power dependence of $Q_{\rm int}$ for all
five resonators at 0.3 K. At near single-photon occupation numbers where TLS
losses dominate the total dissipation, $Q_{\rm int}$ is similar for all the
devices measured. For high photon occupation numbers, $Q_{\rm int}$ is
limited by radiation losses, which varies from device to device by a factor
of $\sim 3$. We attribute this inconsistency to different residual coupling
to parasitics, deviations from left-right symmetry in the prongs of the
tuning fork due to fabrication
tolerances, etc.

To determine the tuning speed of the resonator we followed the same
methodology as in Ref. \citet{sandberg2008tuning}. Figure \ref{Fig 5}
presents the tuning rate characterization measurements. The resonator is
first excited at its unbiased resonance frequency $f_{0}=6.15$ GHz. Then a
rectangular current tuning pulse is applied to detune the resonator to
$f_{0}-\delta f$. With a homodyne detection scheme, we measure the output
microwave power as a function of time, presented in Fig. \ref{Fig 5}(b).
Right after the frequency shift, the output signal is a sum of the excitation
power at frequency $f_{0}$ and the power radiated by the resonator at
$f_{0}-\delta f$; the latter decays on a timescale $\tau=Q/\pi f_{0}$ during
which we observe beatings (Fig. \ref{Fig 5}(b)) in the measured transmitted
power. These beatings have a period precisely equal to $1/\delta f$. By
extracting this period we can infer the instantaneous frequency at very short
timescales. In Fig. \ref{Fig 5}(c) we track these beatings down to
$\lesssim20$ ns, thus demonstrating an almost instantaneous detuning of 40
MHz, i.e. in a time more than 1000 times shorter than the photon lifetime. We
would like to stress that the instantaneous frequency shift of 40 MHz could
not be detected in a time shorter than $\sim 1/40$ MHz $=25$ ns, i.e. the
reported $20$ ns time is actually limited by the resolution of the readout
method itself, rather than the internal device response time, that is
expected to be even faster.

\begin{figure}
\includegraphics{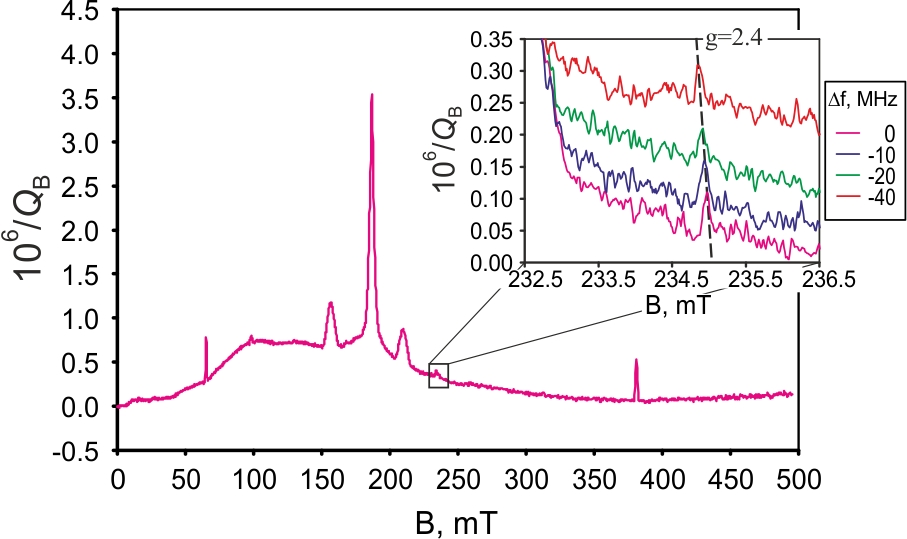}

\caption{\textit{{Tunable resonator as a core of ESR spectrometer}. }Main
plot: ESR spectrum (excessive dissipation as a function of magnetic
field) taken with resonator operating at a base resonance frequency.
The spectrum reveals paramagnetic impurities in sapphire substrate
and the spins of surface absorbants. Inset: a set of ESR spectra taken
with different frequency shifts as indicated on the legend. The slope
of the dashed line corresponds to g-factor 2.4\textit{.}}

\label{Fig 6}
\end{figure}

Previously we have demonstrated that the field resilient fractal resonators
can be used as a core of on-chip Electron Spin Resonance (ESR) spectrometers
\citet{de2017direct,de2018suppression}. Compared to a standard ESR bulk
cavity design, the microwave field volume in a planar resonator is squeezed
up to $~10^{6}$ times resulting in a greatly enhanced spin-to-microwave
coupling. In combination with a high $Q$-factor provided by superconducting
resonators, this translates to exceptional ESR sensitivity, with a minimum
detectable number of spins of a few hundreds \citet{probst2017inductive,
ranjan2020electron}. The microwave field in a planar resonator is naturally
localized around the substrate surface, which provides selective sensitivity
to surface spins and deposited nanoscale clusters
\citet{de2017direct,graaf2012magnetic}. Magnetic field resilience is
instrumental for ESR applications, which generally present a challenge for
superconducting designs. To achieve high-field resilience we reapply the
design ideas previously exploited in \citet{graaf2012magnetic}. Specifically,
we implement the fractalized ground plane design to eliminate the magnetic
flux focusing. The ground plane in both C- and L-coupled designs is split in
such a way that a flux escape path is available for all open (no film) areas,
and superconducting loops are  thus avoided. Finally (and most importantly),
the narrow thin film elements composing a fractal structure effectively expel
vortices. Taking all these precautions, we demonstrate that tunable
resonators can be made equally suitable for surface spins detection. In Fig.
\ref{Fig 6}, we present an ESR spectrum (i.e. magnetic field dependent
dissipation part $1/Q_{\rm B}=1/Q_{\rm int}(B)-1/Q_{\rm int}(0)$) obtained
with the tunable resonator by scanning the magnetic field up to 500 mT. From
our previous studies we know the number of spins constituting the main ESR
peak centered at 180 mT \citet{graaf2012magnetic}. As the area under the peak
scales roughly in proportion with the number of constituent spins, the number
of spins contributing to a peak positioned at 235 mT (Fig. \ref{Fig 6}-inset)
is estimated to be $~3000$. Remarkably, even for such a small number of spins
the signal-to-noise ratio in Fig. \ref{Fig 6}-inset is still quite decent.
The frequency tuning option provides ESR spectrometer with an additional
functionality: by varying the probe frequency by $\Delta f_{0}$ and measuring
the corresponding resonance peak shift $\Delta B$ one can directly assess an
equivalent g-factor $g=h\Delta f_{0}/(\mu_{B}\Delta B)=2.4$ for the peak in
the inset of Fig. \ref{Fig 6}.

\section{{\large{}Summary}}

In summary, we have presented a superconducting microwave resonator design
which allows for frequency tuning without substantial compromise of the
resonator quality factor. We demonstrate that the internal Q-factor holds the
value above $10^{5}$ down to the single photon limit over the entire tuning
range up to 200 MHz. The dominant loss mechanism is residual radiation
through the current bias control lines which could be further suppressed with
low-pass filters. We demonstrated that a full-scale frequency shift can be
performed in a time shorter than a $10^{-3}$ fraction of the photon lifetime.

The quasi-fractal resonator design allows for operation in high magnetic
fields. Here we demonstrated the magnetic field resilience up to 0.5 T. We
argue that by shrinking the minimum design features in the design down to
$\sim 100$ nm or by introducing high-density perforation of the
superconducting film the resonators can be made operational in magnetic
fields above 5 T, as previously demonstrated in \citet{kroll2019magnetic,
samkharadze2016high}.

When the resonator is used for ESR spectrometry, the frequency tuning allows
for determination of the effective g-factor of the spins. Owing to a high
quality factor of our resonators, the g-factor can be reliably measured on
spin ensembles as small as $\sim~1000$ spins, which is close to state of the
art in the field \citet{ranjan2020electron}. In perspective, frequency
tunable resonators pave the way for the implementation of advanced ESR
techniques with on-chip devices, such as pulsed ESR and ELDOR-NMR
\citet{cox2017eldor}. Last, but not least, we foresee many potential
applications for fast tunable high-$Q$ resonators in the rapidly progressing
field of quantum computing.

\section*{{\large{}Acknowledgements}}

The work was jointly supported by the UK department of Business, Energy and
Industrial Strategy (BEIS), the EU Horizon 2020 research and innovation
programme (grant agreement 766714/HiTIMe), the Swedish Research Council (VR)
(grant agreements 2016-04828 and 2019-05480), EU H2020 European Microkelvin
Platform (grant agreement 824109), and Chalmers Area of Advance NANO/2018.

\bibliographystyle{apsrev4-2}
\bibliography{HighQf_tuneres.bbl}

\end{document}